\documentclass{rspublic}

\usepackage[dvips]{graphicx}

\begin{document}

\title[Frictional dissipation at a non-sliding interface]{Elasticity and onset of frictional dissipation at a 
non-sliding multicontact interface}

\author[L. Bureau, C. Caroli and T. Baumberger ]{L. Bureau\footnote{Corresponding author. e-mail: 
lionel.bureau@college-de-france.fr. Present address: 
LPMC, Coll{\`e}ge de France, 11 place Marcelin Berthelot, 75231
Paris cedex 05.},  C. Caroli and T. Baumberger}

\affiliation{Groupe de Physique des Solides, 2 place Jussieu, 75251 
Paris cedex 05, France}

\label{firstpage}

\maketitle

\begin{abstract}{contact stiffness, frictional dissipation, Hertz-Mindlin contacts}
We measure the elastic and dissipative responses of a multicontact interface, formed between the rough surfaces 
of two 
contacting macroscopic solids, submitted to a biased oscillating shear force. We  evidence that beyond a linear
viscoelastic regime, observed at low shear amplitude, the interface response exhibits a dissipative component which
corresponds to the onset of frictional dissipation. The latter regime exists whereas the tangential force applied, 
far from the nominal static threshold, does not provoke any sliding. This result, akin to that of Mindlin for a single
contact, leads us to extend his model of `microslip' to the case of an interface composed of multiple 
microcontacts. While describing satisfactorily the elastic response, the model fails to account 
quantitatively for the observed energy dissipation, which,
we believe, results from the fact that the key assumption of local Coulomb friction in Mindlin's model is not
legitimate at the sub-micrometer scale of the microslip zones within microcontacts between surface asperities.
\end{abstract}

\section{Introduction}
\label{sec:intro}

The frictional 
response of the contact between two macroscopic solids submitted to a 
shear force is commonly described, in the framework of Amontons-Coulomb's law, 
in terms of:
(i) a static force threshold, $F_{s}$, below which no relative displacement is 
supposed to occur, and (ii)
a dynamic friction coefficient defined when stationary sliding 
is established.

However, it is known that frictional dissipation in 
mechanical contacts starts to build up for shear forces {\it lower} than the nominal 
static threshold. This behaviour, which is important in mechanical engineering, for instance in problems
 of fretting
 or damping in structural joints (see {\it e.g.} Goodman 1959, 
Olofsson 1995), also presents a fundamental interest
related to the understanding of the microscopic mechanisms responsible for macroscopic friction.
This issue was first extensively studied by Mindlin 
{\it et al.} (1951), Johnson (1955), 
Courtney-Pratt \& Eisner (1956) and Goodman \& Brown (1962), who evidenced that the 
displacement response 
of Hertzian macroscopic contacts submitted to an oscillating 
tangential force of amplitude $F\ll F_{s}$ exhibited a hysteresis loop 
attributable to an interfacial dissipative process. 
In order to account for this energy loss, Mindlin {\it et al.} (1951) proposed a model
of 'microslip' within the contact zone, based on the following description
(Cattaneo 1938): wherever in the contact zone the tangential ($\tau$) and
normal ($\sigma$) local stresses obey $\tau<\mu \sigma$ (with $\mu$ the friction
coefficient) no relative tangential displacement occurs. Otherwise, shear
provokes slip so that, in the corresponding region, the stresses satisfy
{\it locally} a Coulomb friction law: $\tau=\mu \sigma$. It is then predicted
that the sheared contact is composed of a circular non-sliding zone
surrounded by an annular slipping region whose width increases with the
applied shear force.

Still, Mindlin's description of incipient sliding raises a question: 
down to which length scale is
such an assumption of {\em local} friction valid ? 
Indeed, macroscopic solids generally exhibit
rough surfaces which, when brought into contact, form a {\em 
multicontact interface} (MCI), {\it i.e.} an interface composed of a 
dilute set of microcontacts between asperities. When addressing the 
problem of friction at such an interface, one may therefore wonder whether 
the use of a local friction law is legitimate when dealing with the
micrometer-sized contacts between surface asperities. This in turn
raises the question of the spatial scale of the elementary dissipative events 
responsible for solid friction, a cutoff 
length below which a local friction law cannot be meaningful.

Recently, 
experiments performed on such MCIs showed that (Berthoud \& Baumberger 1998):
\bi 
\item for tangential forces $F\ll F_{s}$, the pinned interface 
responds {\it elastically}, {\it via} the reversible deformation of the 
load-bearing asperities. In this regime, the shear stiffness $\kappa$ 
of the interface can be measured and is well 
accounted for within the framework of Greenwood's model of contact 
between rough surfaces (Greenwood \& Williamson 1966).
\item for $F\lesssim F_{s}$, a creeplike irreversible sliding of the 
solids occurs.
\ei

\begin{figure}[htbp]
$$
\includegraphics{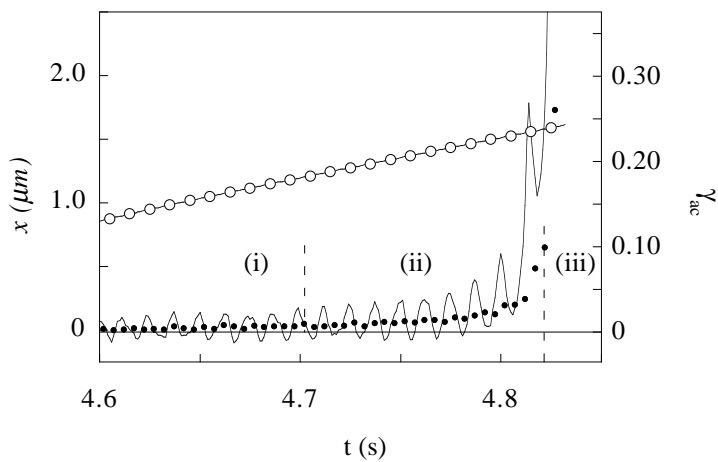}
$$
\caption{Time-plot of the instantaneous (line) and averaged ($\bullet$) displacement response of the slider to a
 biased
oscillating shear force of increasing amplitude. On the right scale is plotted the reduced shear amplitude ($\circ$) 
$\gamma_{ac}(t)=F_{ac}(t)/W$, where $W$ is the normal load. The biased used for this experiment is $\gamma_{dc}
=F_{dc}/W=0.36$ and the frequency of oscillation $f=80$ Hz.}
\label{fig:intro}
\end{figure}

A detailed study of this regime of incipient sliding, below the nominal 
static threshold, should therefore 
provide information about the physical processes underlying 
frictional dissipation. In order to achieve the force control and the 
displacement resolution required to perform such a study, we have
developed an 
experimental setup which allows to probe the response of a 
multicontact interface to a biased oscillating shear force, of 
 amplitude 
$F_{ac}$ about a finite mean value $F_{dc}$, while the maximum force applied 
$F_{dc}+F_{ac}=F_{max}\lesssim F_{s}$ (Baumberger {\it et al.} 1998). We thus showed that the 
displacement response of a macroscopic slider submitted to such a harmonic 
tangential force exhibited three different regimes, depending on 
the amplitude $F_{ac}$ (Bureau {\it et al.} 2001). These are 
illustrated on figure \ref{fig:intro}, which shows the response to a shear force modulation of
slowly increasing amplitude: in region (i), at small $F_{ac}$, the center of 
mass of the slider 
oscillates about a constant average position, which means that no irreversible 
sliding occurs; in region (ii), corresponding to higher shear 
amplitudes, the slider enters a creeplike regime where the 
slipped distance increases continuously, up to a final regime of abruptly
accelerating motion (region (iii)).

In the present article, we report an experimental study which focuses on the 
first regime of small shear amplitudes, in which the interface is pinned, {\it i.e.} oscillates about
a fixed position.
In section \ref{sec:exp}, we briefly present the principle of operation of the experimental 
setup. By measuring the in-phase and out-of-phase components of the displacement response of a
MCI submitted to a harmonic shear force, we can probe accurately
both the elastic and dissipative responses of the interface (section \ref{sec:results}). We 
evidence that:
\bi
\item the shear 
stiffness evolves with the `age' of the interface, {\it i.e.} with the time elapsed since the
solids were brought into contact, in agreement with the creep ageing of the load-bearing 
asperities already identified for such MCIs, which results in the slow logarithmic increase of
microcontact radii (Dieterich \& Kilgore 1994, Berthoud {\it et al.} 1999a),
\item frictional dissipation appears at shear forces well below the nominal static threshold,
while no gross sliding is detected yet.
\ei
This latter point leads us to propose an extension of Mindlin's model to the case of a multicontact interface, 
within the framework of Greenwood's description for the contact of rough surfaces. We find that such a model 
of microslip within the contacts between asperities largely overestimates the energy dissipation, which we believe 
points out that a local description of friction is no longer valid at the sub-micrometric scale of the slip zones 
within the
microcontacts forming multicontact interfaces. 

\section{Experiments}
\label{sec:exp}

\subsection{Samples}
\label{subsec:samples}

The multicontact interface studied here is formed between a slider and a track of 
commercial grade poly(methylmethacrylate) (PMMA). PMMA is a glassy polymer at room temperature 
($T_{g}\simeq~120^{\circ}$C),
of dynamic shear modulus $G'= 2$ GPa and loss angle $\delta$ such that $\tan \delta = 0.1$ at 100 Hz 
(Ferry 1980), and of quasi-static Young modulus $E\simeq 3$~GPa, Poisson ratio $\nu=0.44$ and  hardness 
$H\simeq 300$~MPa (Berthoud {\it et al.} 1999b).

The nominally flat surfaces of the slider (20$\times$20 mm$^{2}$) and of the track (25$\times$30 mm$^{2}$) 
are lapped with an abrasive aqueous suspension of SiC powder (mean particle size 23 $\mu$m). This
leads to a rms roughness of the samples $R_{q}= 1.3$ $\mu$m, as previously characterized (Berthoud \& Baumberger
1998). 

Using Greenwood's result (Greenwood \& Williamson 1966), assuming that: 
(i) the summit heights of asperities follow 
an exponential distribution
of width $s=1.3\, \mu$m, and
(ii) their mean radius of curvature $\beta=20\, \mu$m (taken, as a conservative 
value, of
the order of magnitude of the abrasive particle size), 
one can estimate\footnote{This estimation assumes that contacting asperities are deformed elastically. Actually, 
Greenwood's plasticity index 
$\psi=(E/H)\sqrt{s/\beta}\simeq 2.5$ for our 
surfaces, which indicates that most contacting asperities have started to yield plastically. However, 
since $\psi$ is still 
of order unity, the set of asperities is in fact in an elastic-plastic state of deformation (Berthoud {\it et al.} 
1999b), far from the fully plastic limit. We therefore use, for our rough estimate of $N$, the expression corresponding
to the elastic limit (evaluating $N$ in the fully plastic limit would lead to a number of microcontacts $\psi$ times
larger). }, under the normal load 
$W\simeq 2$ N, the number $N$ of microcontacts: $N\simeq W/(\sqrt{\pi s\beta}E^{*}s)
\simeq 100$ (where $E^{*}=E/[2(1-\nu^2)]$), and their mean radius $\bar{a}\simeq \sqrt{s\beta}\simeq 5\, \mu$m.

\subsection{Experimental setup: an inertial tribometer}

\begin{figure}[htbp]
$$
\includegraphics[width=6cm]{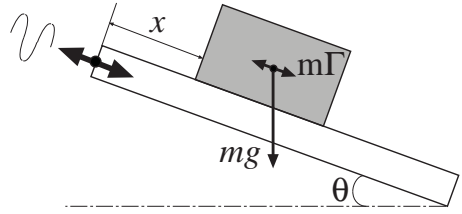}
$$
\caption{Principle of operation of the inertial tribometer: the slider on the inclined track is submitted to 
a constant tangential force $F_{dc}=mg\sin\theta$ on which is superimposed a harmonic shear force of amplitude
$m\Gamma$, where $\Gamma$ is the imposed acceleration amplitude of the track. We measure the displacement $x$ of the
 slider with respect to the track.}
\label{fig:exp1}
\end{figure}

The experimental setup, extensively described elsewhere (Baumberger {\it et al.} 1998) is based 
on the following principle. The track, on which the slider sits under its own weight $mg$, is first 
inclined at a given angle $\theta$ such that the ratio of the tangential ($F_{dc}$) to normal ($W$) load 
$F_{dc}/W=\tan \theta \ll \mu_{s}$. A harmonic motion, of controlled amplitude and frequency, is then imposed
to the track, which results in an inertial oscillating shear force acting on the center of mass of the 
slider (see figure \ref{fig:exp1}). In order to avoid torques and tilt motion, the slider sample
is clamped in a metal part specially 
designed such that the center of mass of the slider is located in the plane of the interface. 

The harmonic  shear force on 
the interface thus reads, in the low frequency limit where the slider responds quasi-statically (see 
Baumberger {\it et al.} 
1998), $F(t)=F_{ac}\cos(\omega t)$ with $F_{ac}=m\Gamma$, where $\Gamma$ is the imposed 
acceleration amplitude of the track. We define $\gamma_{dc}=F_{dc}/W$ and 
$\gamma_{ac}=F_{ac}/W$. These control parameters can be set in the ranges 
$0\leq\gamma_{dc}\leq 0.58$ and $0\leq\gamma_{ac}\leq 0.6$. The frequency of the
oscillating tangential force can be chosen between 15 and 100 Hz, this upper limit ensuring the quasi-static 
condition for the slider motion.

We measure, in response to this excitation, the displacement $x(t)$ of the slider relatively to the track, 
by means of a capacitive gauge. Its signal is sent to a lock-in amplifier, which allows to detect, within 1 nm, the 
in-phase ($x_{0}$) and out-of-phase ($x_{90}$) components of the displacement with respect to the harmonic
input. We thus have access to the elastic and dissipative responses of the multicontact interface.

\subsection{Reproducibility}

The scattering of the results thus obtained depends crucially on the way the interface is prepared. We have
tested two different protocols:

(i)after the slider is put into contact with the inclined track, a time $t_{w}$ is waited during which the 
interface is left under constant normal and tangential loads. At the end of this waiting time, $\gamma_{ac}$ 
is turned on for a time $\ll t_{w}$ during which we measure
$x_{0}$ and $x_{90}$. When performing several such experiments in the same
nominal conditions, the relative dispersion  observed is on the 
order of 25\%.

(ii) the second protocol consists, as soon as the interface has been created, in applying to the slider a large 
amplitude harmonic shear force in
order to make it slide a few micrometers in the direction of $F_{dc}$. The oscillating shear force is then 
suddenly stopped and a time $t_{w}$ is waited, after which measurements are performed. The dispersion
observed when using this second protocol is reduced to 11\%. We will therefore present, in the rest of the
 paper, results that have been obtained on interfaces prepared this way.     

\begin{figure}[h!]
$$
\includegraphics[width=10cm]{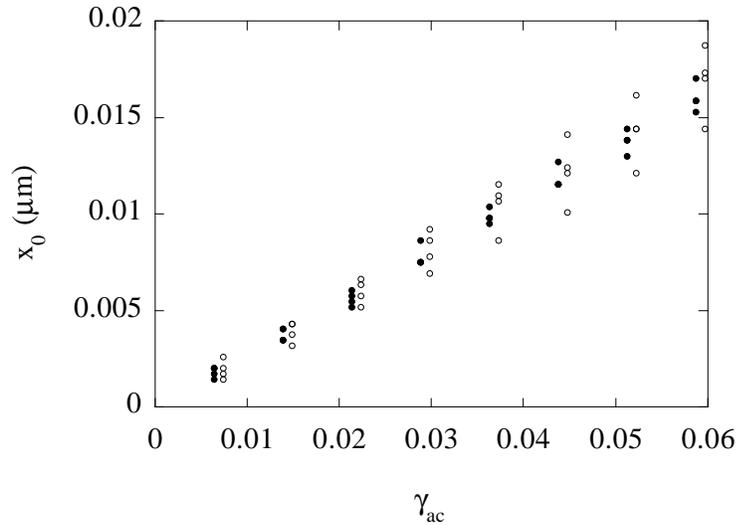}
$$
\caption{Influence of the protocol of interface preparation on the reproducibility. In-phase response $x_{0}
(\gamma_{ac})$ obtained with protocol (i) ($\circ$), and with protocol (ii) ($\bullet$). Four sets of measurements
are plotted, performed
at $f=60$ Hz, with $\gamma_{dc}=0.27$ and after $t_{w}=300$~s.}
\label{fig:reprod}
\end{figure}

This effect of the interface preparation on the reproducibility is illustrated on figure 
\ref{fig:reprod}.

To understand, at least qualitatively, the origin of this effect, note
the following. When the slider is put into contact with the inclined
track, the only mechanical condition which constrains the state of
the MCI is that the {\it total} tangential force $F=W\tan \theta$.
This is obviously insufficient to define uniquely the distribution of
shear forces on the various microcontacts. Hence, a huge number of local
configurations are possible: the interface is a highly multistable
system. On the contrary, sliding produces a reproducible distribution
which is maintained during the elastic recoil following a stop (Caroli \& Nozi{\`e}res 1996, 
de Gennes 1997).
Preparing an interface by interruption of sliding can therefore be
expected to improve the degree of reproducibility.
The remaining scattering of 11\% is indeed in 
agreement with the statistical dispersion $\sim N^{-1/2}\sim 10$\% due to the finite number of microcontacts 
estimated above.

\section{Results}
\label{sec:results}

\begin{figure}[htbp]
$$
\includegraphics[width=10cm]{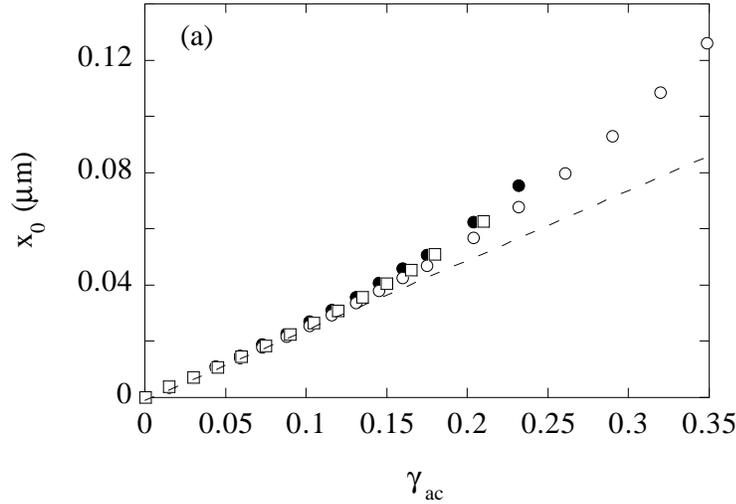}
$$
\caption{In-phase displacement amplitude $x_{0}$ as a function of $\gamma_{ac}$, at $f=80$ Hz and after 
$t_{w}=600$ s. ($\bullet$): $\gamma_{dc}=0$, ($\circ$): $\gamma_{dc}=0.09$, ({\tiny$\square$}): $\gamma_{dc}=0.27$.
}
\label{fig:x0type}
\end{figure}

\begin{figure}[htbp]
$$
\includegraphics[width=10cm]{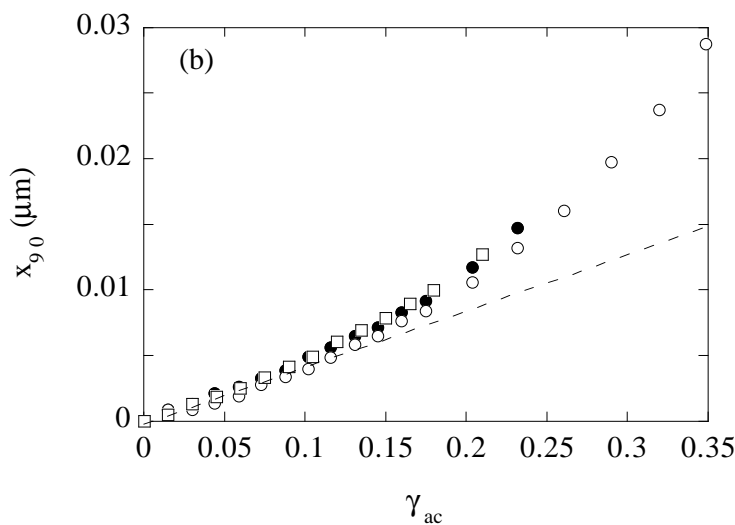}
$$
\caption{Out-of-phase displacement amplitude $x_{90}$ as a function of $\gamma_{ac}$, at $f=80$ Hz and after 
$t_{w}=600$ s. ($\bullet$): $\gamma_{dc}=0$, ($\circ$): $\gamma_{dc}=0.09$, ({\tiny$\square$}): $\gamma_{dc}=0.27$}
\label{fig:x90type}
\end{figure}

We present on figures \ref{fig:x0type} and \ref{fig:x90type} typical results obtained for the in-phase ($x_{0}$)
 and out-of-phase ($x_{90}$) components as a function of the reduced shear force amplitude $\gamma_{ac}$. We do not
observe any dependence of the elastic or dissipative response on the level of average tangential force 
$\gamma_{dc}$ ({\it i.e.} on the angle of inclination of the track, up to $\gamma_{dc}\simeq 0.36$), and do not note
 any frequency dependence in the explored range 15--100 Hz. 

Both $x_{0}(\gamma_{ac})$ and $x_{90}(\gamma_{ac})$ exhibit a linear regime at low shear amplitude, up to 
$\gamma_{ac}\simeq 0.1$ above which the interface responds non-linearly. For all the results presented in this 
article, in the range of $\gamma_{ac}$ explored, {\em no sliding is 
detected} within the experimental resolution $\lesssim 20$~nm.

\subsection{Linear response}

In the linear regime, we measure the interfacial shear stiffness, which reads $\kappa=\gamma_{ac}W/x_{0}$.
Moreover, we know that, for a multicontact interface, this stiffness varies 
proportionally to the normal load: $\kappa=W/\lambda$, where $\lambda$ is an `elastic length' which lies in the
 micrometer range (Berthoud \& Baumberger 1998, see also section \ref{sec:discuss}\ref{subsec:elast} for more 
details). The slope of 
$x_{0}(\gamma_{ac})$ is thus $\lambda=\text{d}x_{0}/\text{d}\gamma_{ac}$.
We measure, after a waiting time $t_{w}=300$~s, $\lambda=0.26\pm 0.015\, \mu$m.

\begin{figure}[htbp]
$$
\includegraphics[width=10cm]{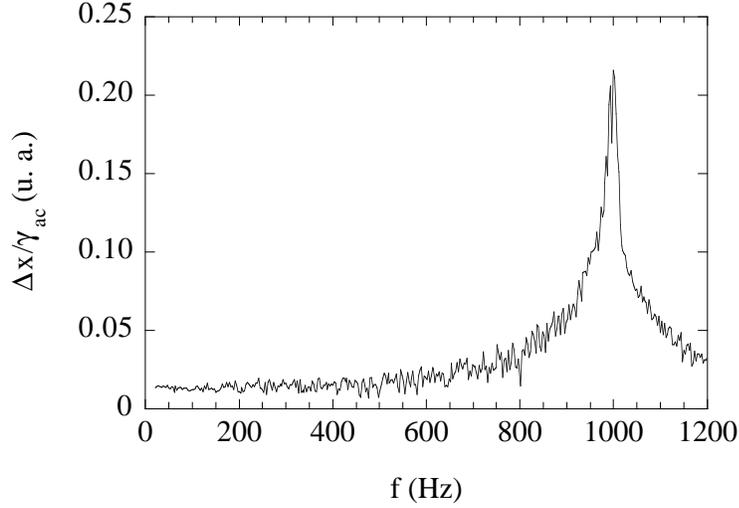}
$$
\caption{Spectral response: the slider is submitted to an acceleration of constant amplitude and variable frequency.
The ratio of the displacement amplitude $\Delta x$ to the acceleration amplitude is plotted, in arbitrary units, as 
a function of $f$. The bias $\gamma_{dc}=0$.}
\label{fig:spectre}
\end{figure}

This length can also be determined by analyzing the frequency response of the system. Indeed, we expect that the 
response of the slider, of mass $m$, sitting on the elastic foundation of stiffness $\kappa$ formed by the set of 
load-bearing asperities, exhibits a resonance at a circular frequency $\omega_{0}=\sqrt{\kappa/m}=\sqrt{g/\lambda}$
(with $W=mg$), as already observed by Sherif \& Kossa (1991).
When looking at the spectral response shown on figure \ref{fig:spectre}, one clearly identifies a peak at $f_{0}=
1000$~Hz which corresponds to that resonance and leads to a value of the elastic length $\lambda=0.25\, \mu$m, in
excellent agreement with the value reported above.

\bigskip

Let us now try to identify the origin of the dissipation in this regime of very small shear
amplitudes. We note that the inverse of the quality factor
$1/Q=0.03$ of the resonance of figure \ref{fig:spectre} is comparable with the
tangent of the loss angle $\tan \delta \simeq 0.05$ of bulk PMMA at
1kHz and $T=300$ K (Ferry 1980).
Besides, the ratio $x_{90}/x_{0}=0.18$ is constant and on the order of 
$\tan \delta
\approx 0.1$ at $f\lesssim 100$~Hz.
This leads us to attribute the observed dissipation to the viscoelastic
losses {\em within the bodies of contacting asperities} -- more precisely,
for each of them, within the volume of order roughly $\bar{a}^{3}$
(with $\bar{a}$ the mean contact radius) in which stresses concentrate.
The discrepancy observed between $x_{90}/x_{0}$ and the loss angle measured on 
bulk samples can be assigned to the fact that these micrometric volumes lie
in the interfacial region, and hence most probably present mechanical
properties slightly different from those of the bulk. Indeed, our
method for surface abrasion (see \S \ref{sec:exp}\ref{subsec:samples}) makes use of water, known as
a plasticizer of PMMA.

\subsection{Ageing}

\begin{figure}[htbp]
$$
\includegraphics[width=10cm]{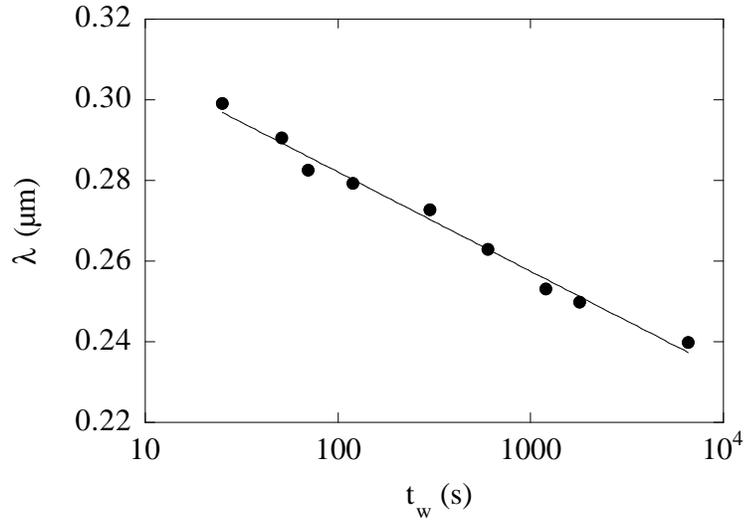}
$$
\caption{Decrease of the elastic length $\lambda$ with the waiting time $t_{w}$. $f=80$ Hz, $\gamma_{dc}=0.18$, 
$\gamma_{ac}=0.04$.}
\label{fig:ageing}
\end{figure}

When performing measurements in the linear regime at various waiting times $t_{w}$, we note that the interfacial
shear stiffness, or equivalently the elastic length $\lambda$, evolves slowly with $t_{w}$: the longer $t_{w}$,
the lower $\lambda$, {\it i.e.} the higher the stiffness. This ageing of the interface is illustrated on figure
\ref{fig:ageing}. The elastic length decreases quasi-logarithmically with
the age of the interface:

\begin{equation}
\lambda(t_{w})=\lambda_{0}-\beta_{\lambda}\ln(t_{w}/t_{0})
\label{eq:lambdadet}
\end{equation}
with the reference time $t_{0}=1$~s, $\lambda_{0}=0.33$ and the logarithmic slope $\beta_{\lambda}=1.07
\times 10^{-2}$.

\subsection{Non-linear regime: interfacial dissipation}

At shear amplitudes $\gamma_{ac}>0.1$, the in-phase and out-of-phase components of the response increase 
non-linearly. In this regime, while the bulk response of the asperities remains linear (one can estimate a mean 
shear strain as the ratio of $x_{0}$ to the mean contact size $\bar{a}$, which stays lower than 2\%), 
the ratio $x_{90}/x_{0}$ is not constant anymore, hence the energy loss cannot be attributed to bulk
 viscoelasticity  only.

\begin{figure}[htbp]
$$
\includegraphics[width=10cm]{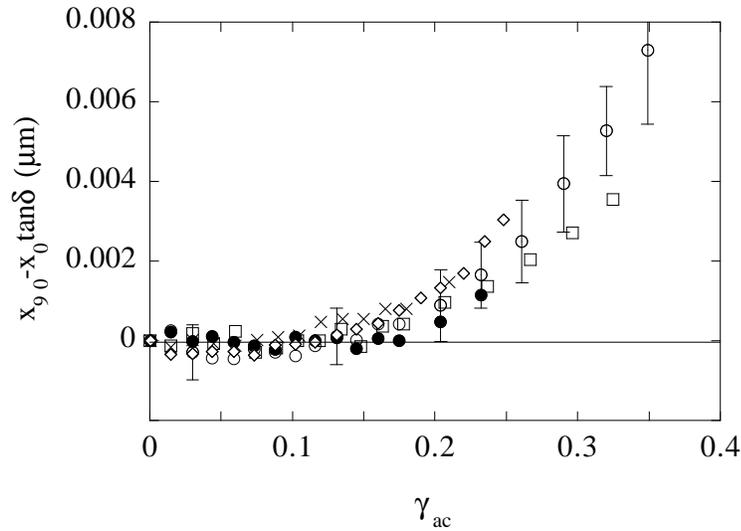}
$$
\caption{Non-viscoelastic component of the dissipative response, $x_{90}-x_{0}\tan\delta$, as a function
of $\gamma_{ac}$. $f=80$ Hz and $t_{w}=600$ s. ($\bullet$): $\gamma_{dc}=0$, ($\circ$): $\gamma_{dc}=0.09$,
({\tiny $\square$}): $\gamma_{dc}=0.18$, ($\times$): $\gamma_{dc}=0.27$, ($\diamond$): $\gamma_{dc}=0.36$.}
\label{fig:dissip}
\end{figure}

We show on figure \ref{fig:dissip} the evolution, with shear amplitude, of the non-viscoelastic part of the 
dissipative response, which we define as $x_{90}-x_{0}\tan\delta$, using for $\tan\delta$ the value
$x_{90}/x_{0}=0.18$ determined in the linear regime.

\section{Discussion}
\label{sec:discuss}

\subsection{Interfacial shear stiffness}
\label{subsec:elast}

Let us consider the case of an infinitesimal shear modulation. In this limit, the shear stiffness of a single contact is 
simply given by the Mindlin expression calculated in the absence of slip, namely $\mathcal{G}
a$, with $a$ the contact radius and $\mathcal{G}
=4G/(2-\nu)$, where $\nu$ and $G$ are respectively the
Poisson ratio and shear modulus of the contacting materials (Mindlin 1949). The shear stiffness of the 
multicontact interface then reads (Berthoud \& Baumberger 1998):
\begin{equation}
\label{eq:kappa}
\kappa\cong N\mathcal{G}\bar{a}
\end{equation}
with $N$ the number of microcontacts and $\bar{a}$ their mean radius (this result is rederived in the Appendix). 
An important feature of
 Greenwood's
description of the contact between rough surfaces is that the number of microcontacts varies linearly with the 
normal load $W$, whereas their mean size is independent of $W$. So, the interfacial stiffness is proportional 
to the load, {\it i.e.} $\kappa=W/\lambda$, with an elastic length 
reading:
\begin{equation}
\label{eq:lambda}
\lambda\cong \frac{H}{\mathcal{G}}\sqrt{s\beta}
\end{equation}
Using for the PMMA shear modulus its quasi-static value $G\simeq 1$ GPa, Berthoud \& Baumberger found from equation
\ref{eq:lambda} an elastic length $\lambda\simeq 1 \, \mu$m, in agreement with their quasi-static measurements of
the stiffness during loading-unloading cycles. 

Our results, however, lead to values of $\lambda\simeq 0.25 \, \mu$m much smaller than those previously reported. 
This marked difference may
have two distinct origins:
\bi
\item First, we clearly see from our experiments that the linear regime of interfacial response corresponds to 
elastic displacements of the slider of at most 20 nm (see figure \ref{fig:x0type}). Such a resolution could not be
achieved in the previous quasi-static experiments, and the stiffness measured in that study was most probably 
underestimated, due to non-linear effects, which thus led to overestimated values of $\lambda$.
\item Moreover, note from expression \ref{eq:lambda} that the elastic length is inversely proportional to the
shear modulus. We thus expect the elastic response of the multicontact interface to be governed, in our experiments,
by the 
{\it dynamic modulus} at the excitation frequency, {\it i.e.} $G'\simeq 2$ GPa (see \S 
\ref{sec:exp}\ref{subsec:samples}).
\ei
With this latter value for the dynamic modulus, along with those for the hardness and the surface characteristics 
given in section \ref{sec:exp}\ref{subsec:samples}, we obtain $\lambda\simeq 0.39 \, \mu$m. The elastic length 
that we estimate from the elastic properties of PMMA at the 
excitation frequency is therefore in good agreement with our experimental results.

\subsection{Ageing}

We now address the question of the time-dependence of the elastic length. It is well established, since the 
pioneer work of Bowden \& 
Tabor (1950), that the real area of contact ($\Sigma_{r}$) between rough surfaces is in general a very small fraction
of the nominal area. The normal stresses on the load-bearing asperities are thus on the order of the 
yield stress of the contacting materials, which results in bulk plastic creep of these asperities. As a consequence,
the real area of contact slowly increases with the `age' of the interface, {\it i.e.} with the time of contact 
between asperities, as unambiguously evidenced by Dieterich \& Kilgore (1994). From an extensive study of the
mechanical properties of polymer glasses, Berthoud {\it et al.} (1999a) have shown that creep of the load-bearing 
asperities results in a quasi-logarithmic growth of $\Sigma_{r}$:
\begin{equation}
\label{eq:creep1}
\Sigma_{r}\cong \Sigma_{r0}\left[1+m\ln\left(1+\frac{t_{w}}{t_{c}}\right)\right]
\end{equation}
with $m$ and $t_{c}$ two material parameters that can be identified from bulk mechanical tests. For PMMA at room
temperature, $m\in [0.04-0.05]$, and a higher bound\footnote{The cut-off time $t_{c}$ could not be determined 
accurately at
room temperature, and was inferred from 
the velocity dependence of the friction force (Baumberger {\it et al.} 1999, Bureau 2002).} for the cut-off time is 
$t_{c}<5\times 10^{-3}$ s.

From expression \ref{eq:kappa} of the shear stiffness $\kappa=N\mathcal{G}\bar{a}=W/\lambda$, and with
the real area of contact $\Sigma_{r}\cong N\bar{a}^{2}$, we find that the elastic length reads:
\begin{equation}
\label{eq:lambdadet2}
\lambda\cong \frac{W}{\mathcal{G}\sqrt{N\Sigma_{r0}\left[1+m\ln(1+t_{w}/t_{c}\right]}}
\end{equation}
To first order in $m$, this expression reads:
\begin{equation}
\label{eq:lambdadet3}
\lambda\cong \underbrace{\frac{W}{\mathcal{G}\sqrt{N\Sigma_{0}}}}_{\lambda_{0}}\left[1-\frac{m}{2}\ln
\left(1+\frac{t_{w}}{t_{c}}\right)\right]
\end{equation}

We thus expect the elastic length to decrease quasi-logarithmically with $t_{w}$, with a logarithmic slope of $m/2$.
Indeed, when fitting the data of figure \ref{fig:ageing} with an expression of the form $\lambda=\lambda_{0}
\left[1+\xi\ln(1+t_{w}/t_{c})\right]$, leaving $\xi$ and $t_{c}$ as free parameters, the best fit is obtained for
$\xi=0.024$ and $t_{c}=10^{-3}$~s, these values being in full agreement with the expected ones.
The observed dependence of the interfacial shear stiffness on the waiting time $t_{w}$ thus results from the creep
ageing of the microcontacts. 

It is interesting to note that up to now, this mechanism of interfacial ageing has always been characterized through
the time-dependence of the static friction threshold (see Berthoud {\it et al.} 1999a, and references therein), which 
is a `destructive' method 
in the sense that the set
of load-bearing asperities is renewed by sliding when the measurement is performed. On the contrary, our 
low-amplitude oscillating shear experiments provide a way to probe accurately the slowly evolving viscoelastic
response of a given set of microcontacts, without any macroscopic sliding at the interface. 

\subsection{Non-linear elasticity and energy dissipation: extension of Mindlin's model}

Figures \ref{fig:x0type}
and \ref{fig:x90type} clearly show that, for $\gamma_{ac}\gtrsim
0.1$, the interfacial elastic response becomes
non-linear, while no gross sliding is observable. In the same regime,
a non-linear dissipative response develops on top of the linear term
attributable to bulk viscoelasticity (figure 8). This contribution
must therefore be considered as resulting from interfacial
dissipation proper.

The decrease of the `local' interfacial stiffness,
$d\gamma_{ac}/dx_{0}$, with increasing shear amplitude may be
interpreted qualitatively as follows. The diameters of the
microcontacts which form the interface are statistically distributed
about the average value $\bar{a}$. A finite shear necessarily leads to
destroying the smaller ones. The larger the shear amplitude, the
smaller the number of microcontacts which are still able to sustain
the stress, hence a decreasing stiffness.

In order to describe quantitatively this regime, we now extend
Mindlin's description to the case of a multicontact interface as
follows:

(i) The contact between the two rough surfaces is described {\it \`a
la} Greenwood, with the assumption of an exponential distribution of
summit heights, and of elastic deformation of asperities. We believe the latter 
assumption to be inessential: indeed, we saw in \S\ref{sec:exp}\ref{subsec:samples}
that though contacting asperities are in an elastic-plastic state, the value of the plasticity index $\psi$ is of
order unity, which suggests that the normal stress profile in microcontacts is still close to the Hertz profile 
(Johnson 1985).

(ii) Mindlin's results give, for a given microcontact, the expression of the
tangential force
associated with a remote shear displacement $x$. This displacement,
equal to that of the center of mass of the slider, is common to all
microcontacts.

(iii) For any finite $x$, there always exists a set of small
microcontacts which are completely slipping. For these, the tangential
force is saturated at its constant maximum value $\mu w$, where $w$ is
the normal load on the microcontact.

(iv) The tangential force on the slider is simply the sum of those on
the various microcontacts.

The detailed calculation is performed in the Appendix.
It yields the following results:
\begin{equation}
    \label{eq:x0}
    x_{0}=2\mu \lambda \left[\frac{\gamma_{ac}}{2\mu}+
    \left(\frac{\gamma_{ac}}{2\mu}\right)^{2}+
    \frac{5}{4}\left(\frac{\gamma_{ac}}{2\mu}\right)^{3}+
    \mathcal{O}\left(\frac{\gamma_{ac}}{2\mu}\right)^{4}\right]
\end{equation}
\begin{equation}
    \label{eq:x90}
    x_{90}=\frac{4\mu \lambda}{\pi}\left[
    \left(1-\frac{2\mu}{\gamma_{ac}}\right)\ln\left(1-\frac{\gamma_{ac}}{\mu}
    \right)-2\right]=\frac{2\mu\lambda}{3\pi}\left[\left(\frac{\gamma_{ac}}{\mu}\right)^{2}+
\left(\frac{\gamma_{ac}}{\mu}\right)^{3}+
 \mathcal{O}\left(\frac{\gamma_{ac}}{\mu}\right)^{4}\right]
\end{equation}
with $\lambda$ the elastic length defined above. The local
friction coefficient of Mindlin's model, $\mu$, is our single fitting parameter.

\begin{figure}[h!]
$$
\includegraphics[width=10cm]{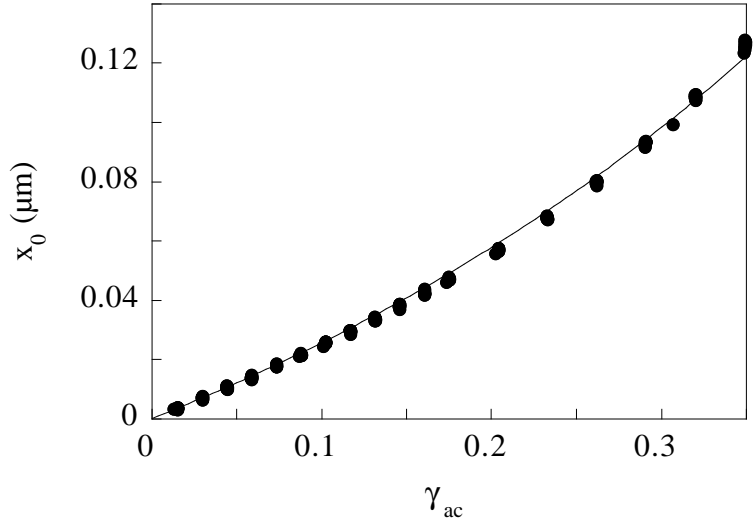}
$$
\caption{($\bullet$): experimental data for $x_{0}(\gamma_{ac})$ at $f=80$~Hz and $\gamma_{dc}=0.09$. (----): 
extended Mindlin's model with $\lambda=0.23\, \mu$m and $\mu=0.49$.}
\label{fig:fitx0}
\end{figure}

Figure \ref{fig:fitx0} shows the best fit thus obtained for the elastic part
of the response $x_{0}$, which is seen to be excellent. It corresponds to
$\mu=0.49$. On the other
hand, from the response to a linear ramp of shear amplitude (see figure
\ref{fig:intro}), we have estimated the (`global') static friction
coefficient $\mu_{s}$ as corresponding to a threshold of accelerated
sliding. We thus find $\mu_{s}=0.59\pm 0.03$ (Bureau {\it et al.} 2001). This can
only be considered as a rough estimate, in view of the arbitrariness
in the definition of the threshold. The agreement therefore appears quite satisfactory.

\medskip

\begin{figure}[h!]
$$
\includegraphics[width=10cm]{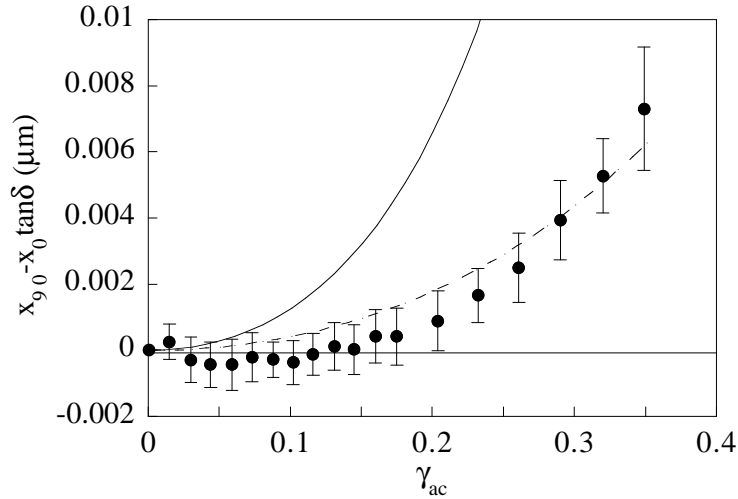}
$$
\caption{($\bullet$): experimental data for $x_{90}(\gamma_{ac})$ at $f=80$~Hz and $\gamma_{dc}=0.09$. (----): 
extended Mindlin's model with $\lambda=0.23\, \mu$m and $\mu=0.49$.
(-- -- --): 
extended Mindlin's model with $\lambda=0.23\, \mu$m and $\mu=1.3$.}
\label{fig:fitx90}
\end{figure}

At this stage, we have determined all the parameters of the model, and are thus in a position to truly check its 
validity by comparing its prediction for $x_{90}(\gamma_{ac})$ with the experimental results. The dissipation 
calculated from equation (\ref{eq:x90}) with $\mu=0.49$ is plotted on figure \ref{fig:fitx90}: it is seen to be much 
larger than the experimental one.
In order to get a decent fit of
the data, we have to use a value of $\mu$ as large as 1.3, {\it i.e.}
much larger than any value of the friction coefficient ever reported
for PMMA. The question therefore arises of what might be the physical
reason for such a discrepancy which affects primarily the dissipative
part of the response.

In Mindlin's model, dissipation results from the slip at the periphery
of the contact. The inner radius of this annulus is, for an average
microcontact, $c=\bar{a}\left[1-f/(\mu w)\right]^{1/3}$. For
$\bar{a}=5\, \mu$m and taking $f/w=0.2$\footnote{We use for the ratio $f/w$ a typical value of the macroscopic
ratio $F/W$. This amounts to assuming that $N$ identical microcontacts
of size $\bar{a}$ bear the same fraction of normal ($w=W/N$) and tangential ($f=F/N$) 
load.}
we find, for $\mu=0.49$ that the width of the annulus $\bar{a}-c=750$~nm.
Over this distance, the shear stress varies from its maximum value to zero at the edge of the contact. We have thus
tacitly assumed that the Coulomb law is valid on a spatial scale much smaller
than this width.

It is now well documented that frictional dissipation results from
the depinning of structural elements located within the adhesive
junctions between load-bearing asperities. The typical size of these
elements is found to be nanometric (Nakatani 2001, Baumberger {\it et al.} 1999, Bureau {\it et al.} 2002).
The friction force, as it is usually defined, is an average
over the dynamics of a large ensemble of such elements. So, a
reasonably meaningful friction coefficient cannot be defined on a
scale smaller than, say, a hundred nanometers. This must be understood
as a cut-off length below which the Mindlin stress profile probably becomes
inaccurate. The above estimate of $\bar{a}-c$ therefore suggests that
multicontact interfaces with micrometric roughness might be out of
the range of quantitative applicability of Mindlin's model.

\appendix{Extension of Mindlin's model}

\subsection{Mindlin's results}

We first recall the results derived by Mindlin for the contact of identical elastic spheres
submitted to a tangential force (Mindlin {\it et al.} 1953, Johnson 1985). When the contact is first loaded from zero,
 the relationship between the remote 
tangential displacement $x$ and
the applied shear force $f$ reads:
\begin{equation}
\label{eq:mind1}
x= \frac{3(2-\nu)\mu w}{8Ga} \left[1-\left(1-\frac{f}{\mu 
    w}\right)^{2/3}\right]
\end{equation}
with $G$ the shear modulus, $\nu$ the Poisson ratio, $a$ the Hertz radius of contact, $w$ the normal load and $\mu$
 the local friction coefficient.

If the shear force is then decreased, after having reached a maximum value $f_{max}$, the displacement $x_{\searrow}$ 
in this unloading phase is:
\begin{equation}
    \label{eq:mind2}
    x_{\searrow}=\frac{3(2-\nu)\mu 
w}{8Ga}\left[2\left(1-\frac{f_{max}-f}{2\mu 
w}\right)^{2/3}-\left(1-\frac{f_{max}}{\mu w}\right)^{2/3}-1\right]
\end{equation}

By symmetry, if the shear force is then reversed from a value $-f_{max}$, the displacement $x_{\nearrow}(f)=-x_{\searrow}
(-f)$.

\subsection{Contact between rough surfaces}

The contact  geometry is that of two rough surfaces of identical rms roughness $\sigma$.
We shall consider, within the framework of Greenwood's model,
the case of contact between a rigid, ideally smooth, reference plane and a composite rough surface whose
 summit heights are distributed exponentially: $\phi(z)=s^{-1}exp(-z/s)$, with $s=\sqrt{2}\sigma$ (Berthoud \& 
Baumberger 1998). The coordinate
 $z$ is normal to the 
mean plane of the random surface.

 The effective elastic modulus of the deformable material is defined as 
$E^{*}=E/[2(1-\nu^{2})]$, and the equivalent radius of curvature at the tip of asperities for the composite surface is 
$R^{*}=R/\sqrt{2}$. 

For a given normal load on the solids, we note $h$ the distance between the mean plane
 of the rough surface and the reference flat. The compression of a contacting asperity of height $z>h$ is thus
$\delta=z-h$ (Greenwood \& Williamson 1966).

\subsection{First loading of the multicontact interface}
\label{subsec:firstload}

We now calculate the interface response when the shear displacement $x$ first increases from 0 to $x_{max}$.

For each microcontact, the Hertz radius $a$ and the normal load $w$ depend on the compression $\delta=z-h$:
\begin{equation}
\label{eq:a1}
a=\sqrt{R^{*}\left(z-h\right)}
\end{equation}
\begin{equation}
\label{eq:w1}
w=\frac{4}{3}\frac{E^{*}}{R^{*}}a^{3}=\frac{2}{3}\frac{E}{(1-\nu^{2})}
\sqrt{R^{*}}\left(z-h\right)^{3/2}
\end{equation}

Plugging (\ref{eq:a1}) and (\ref{eq:w1}) into equation (\ref{eq:mind1}), and using
 $G=E/(2(1+\nu))$, we obtain:
\begin{equation}
    \label{eq:M0}
\left(1-\frac{f}{\mu w}\right)^{2/3}=1-\frac{x s}{\mu \lambda (z-h)}
\end{equation}
where $\lambda=s(2-\nu)/[2(1-\nu)]$. 

We note that the rhs term of equation (\ref{eq:M0}) is $\geq 0$ for:
\begin{equation}
\label{eq:C0}
z-h\geq \frac{xs}{\mu\lambda}
\end{equation}
This means that
 microcontacts whose
compression $z-h$ satisfies condition (\ref{eq:C0}) bear a tangential force
$f\leq\mu w$. Microcontacts such that $z-h=xs/(\mu\lambda)$ are totally sliding, {\it i.e.} $f=\mu w$.

 In the 
following, we will assume that `small' contacts such that $z-h<xs/(\mu\lambda)$ are also sliding and bear a tangential
 force equal to $\mu w$\footnote{Actually, when the interface is sheared, the smallest microcontacts are destroyed and 
replaced by new ones, the contribution of which we cannot calculate. We will see further that the interfacial response
is not significantly affected by this assumption.}.

Hence, for microcontacts such that (\ref{eq:C0}) is verified, equation (\ref{eq:M0}) leads to:
\begin{equation}
    \label{eq:M0bis}
\frac{f}{\mu}=\frac{2E}{3(1-\nu^{2})}\sqrt{R^{*}}\left(z-h\right)^{3/2}\left[1-
\left(1-\frac{x s}{\mu \lambda (z-h)}\right)^{3/2}\right]
\end{equation}

The total tangential force on the system is obtained by integration over the height distribution:
{\setlength\arraycolsep{2pt}
\begin{eqnarray}
    \label{eq:M1}
\frac{F}{\mu} & = & \frac{2E}{3(1-\nu^{2})}\sqrt{R^{*}}N_{0}\left\{
\int_{h_{0}}^{+\infty}
\left[\left(z-h\right)^{3/2}-\left(z-h-\frac{x s}{\mu 
\lambda}\right)^{3/2}\right]\frac{1}{s}e^{-z/s}dz \right. \nonumber \\ 
& & \left. +\int_{h}^{h0}\left(z-h\right)^{3/2}\frac{1}{s}e^{-z/s}dz \right\}
\end{eqnarray}}
where $N_{0}$ is the total number of asperities, and $h_{0}$ is given by condition (\ref{eq:C0}). The first integral 
corresponds to the contribution of microcontacts whose response is given by (\ref{eq:M0bis}), while the second term 
corresponds to totally sliding contacts.

Setting $\xi=(z-h)/s$, and noting that the total normal load on the interface reads, according to 
Greenwood:
\begin{equation}
W=\underbrace{(4/3)E^{*}\sqrt{R^{*}}N_{0}s^{3/2}e^{-h/s}}_{W_{0}}\int_{0}^{+\infty}\xi^{3/2}e^{-\xi}d\xi
\end{equation}
we finally get the following expression for the macroscopic shear force:
\begin{equation}
    F=\mu W\left(1-e^{-\frac{x}{\mu\lambda}}\right)
\label{eq:M3prim}
\end{equation}
Inverting (\ref{eq:M3prim}) yields:
\begin{equation}
    x=-\mu\lambda \ln\left(1-\frac{F}{\mu W}\right)
\label{eq:M3}
\end{equation}

In the limit of small tangential displacements, to lowest order in $x/\mu \lambda \ll 1$,
$F=Wx/\lambda$. We thus find the expression of the interfacial shear stiffness $\kappa=W/\lambda$, where 
$\lambda=s(2-\nu)/[2(1-\nu)]$ is the elastic length. For numerical purposes, we will however not estimate $\lambda$ from
this expression but will rather make use of its experimentally
measured value $\lambda\simeq 0.25\, \mu$m.

Finally, evaluating, from equation (\ref{eq:M1}), the relative contribution of totally sliding microcontacts, we find
that they contribute less than 10\% to the calculated shear force\footnote{This estimate is done using 
$\mu=0.49$, the value which leads to the best fit of $x_{0}(\gamma_{ac})$ (see figure \ref{fig:fitx0}). A 
higher value of $\mu$ would lead to an even weaker contribution of sliding contacts.} while the tangential displacement 
stays
 lower than 60 nm (which corresponds to reduced shear force amplitudes $\gamma_{ac}\lesssim 0.2$, see {\it e.g.} figure
\ref{fig:x0type}). Their contribution is at most 20\% for $x\simeq 100$ nm.
   
\subsection{Unloading from $(F_{max},\; x_{max})$}

Let us now study the case where the tangential displacement of the slider is decreased, after having reached a maximum 
value $x_{max}$, corresponding to a maximum
shear force $F_{max}$.
Two families of microcontacts must then be considered:

\subsubsection{Microcontacts such that $(z-h)<x_{max}s/(\mu \lambda)$}

At the end of the first loading, these microcontacts are totally sliding.
For $x=x_{max}$, the shear force on one of them is $f_{max}=\mu w$, and its response when $x$ is decreased is given 
by (Mindlin {\it et al.} 1953, Johnson 1985):
\begin{equation}
x_{\searrow}=x_{max}-\frac{3(2-\nu)\mu 
w}{4Ga}\left(1-\frac{b^{2}}{a^{2}}\right)
\end{equation}
where $b$ is the inner radius of the corresponding slip zone: $b=a[1/2+f/(2\mu w)]^{1/3}$.
We thus obtain:
\begin{equation}
\label{eq:M4prim}
\left(\frac{1}{2}+\frac{f}{2\mu w}\right)^{2/3}=1-\frac{(x_{max}-x_{\searrow})s}
{2\mu\lambda (z-h)}
\end{equation}
As in \S\ref{subsec:firstload}, this equation yields the following condition: 
\begin{equation}
z-h\geq \frac{(x_{max}-x_{\searrow})s}{2\mu \lambda}
\label{eq:C2}
\end{equation}
Microcontacts whose compression satisfies condition (\ref{eq:C2}) respond according to equation (\ref{eq:M4prim}),
 while those for which $z-h<(x_{max}-x_{\searrow})s/(2\mu \lambda)$ are assumed to be totally sliding and bear a 
tangential force $f=-\mu w$.

For the contacts which are not fully sliding during unloading, equation (\ref{eq:M4prim}) yields:
\begin{equation}
    \label{eq:M4}
\frac{f}{\mu}=\frac{2E}{3(1-\nu^{2})}\sqrt{R^{*}}\left\{2\left[z-h-
\frac{(x_{max}-x_{\searrow})s}{2\mu 
\lambda}\right]^{3/2}-(z-h)^{3/2}\right\}
\end{equation}

\subsubsection{Microcontacts such that $(z-h)\geq x_{max}s/(\mu \lambda)$}

The response of these microcontacts, which were not totally sliding when $x$ reached its maximum value, is given
by (Mindlin {\it et al.} 1953, Johnson 1985):
\begin{equation}
\label{eq:M4bis}
x_{\searrow}=\frac{3(2-\nu)\mu 
w}{8Ga}\left[2\left(1-\frac{f_{max}-f}{2\mu 
w}\right)^{2/3}-\left(1-\frac{f_{max}}{\mu w}\right)^{2/3}-1\right]
\end{equation}
From equation (\ref{eq:M0}) we get:
\begin{equation}
\left(1-\frac{f_{max}}{\mu w}\right)^{2/3}=1-\frac{x_{max}s}{\mu 
\lambda (z-h)}
\end{equation}
which, once plugged into (\ref{eq:M4bis}), leads to the following expression for the shear force on one of these 
microcontacts:
{\setlength\arraycolsep{2pt}
\begin{eqnarray}
    \label{eq:M5}
    \frac{f}{\mu} & = & \frac{2E}{3(1-\nu^{2})}\sqrt{R^{*}}\left\{2\left[z-h-
    \frac{(x_{max}-x_{\searrow})s}{2\mu 
\lambda}\right]^{3/2} \right. \nonumber \\
&& \left. -\left[z-h-\frac{x_{max}s}{\mu 
\lambda}\right]^{3/2}-(z-h)^{3/2}\right\}
\end{eqnarray}}

\subsubsection{Total shear force}

The macroscopic shear force when the interface is unloaded is the sum of the contributions (\ref{eq:M4}),
(\ref{eq:M5}), and of that resulting from fully sliding microcontacts:
{\setlength\arraycolsep{2pt}
\begin{eqnarray}
    \frac{F}{\mu} & = & W_{0}\left\{
\int_{\frac{x_{max}}{\mu\lambda}}^{+\infty}\left[2\left(\xi-\frac{x_{max}-
x_{\searrow}}
{2\mu\lambda}\right)^{3/2}-\left(\xi-\frac{x_{max}}{\mu\lambda}\right)^{3/2}-\xi^{3/2}\right]
e^{-\xi}d\xi \right. \nonumber \\
&&+\int^{\frac{x_{max}}{\mu\lambda}}_{\frac{x_{max}-x_{\searrow}}{2\mu\lambda}}
\left[2\left(\xi-\frac{x_{max}-x_{\searrow}}
{2\mu\lambda}\right)^{3/2}-\xi^{3/2}\right]
e^{-\xi}d\xi \nonumber \\
&&\left. +\int^{\frac{x_{max}-x_{\searrow}}{2\mu\lambda}}_{0}
-\xi^{3/2}e^{-\xi}d\xi \right\}
\end{eqnarray}}
which yields:
\begin{equation}
\label{eq:M6}
\frac{F}{\mu 
W}=2e^{-\frac{x_{max}-x_{\searrow}}{2\mu\lambda}}-e^{-\frac{x_{max}}
{\mu\lambda}}-1
\end{equation}
Setting $\gamma=F/W$ yields:
\begin{equation}
    x_{\searrow}=x_{max}+2\mu \lambda 
    \ln\left(1+\frac{\gamma-\gamma_{max}}{2\mu}\right)
\end{equation}
with $x_{max}$ as given by (\ref{eq:M3}):
\begin{equation}
x_{max}=-\mu \lambda \ln\left(1-\frac{\gamma_{max}}{\mu}\right)
\end{equation}

Finally, by symmetry, the relationship $x(F)$ when the displacement, having reached a minimum value $x_{min}$, is
 reversed up to $x_{max}$, reads:
\begin{equation}
    x_{\nearrow}=x_{min}-2\mu \lambda 
    \ln\left(1-\frac{\gamma-\gamma_{min}}{2\mu}\right)
\end{equation}
with the minimum value of the displacement:
\begin{equation}
x_{min}=x_{max}+2\mu 
\lambda\ln\left(1-\frac{\gamma_{max}-\gamma_{min}}{\mu}\right)
\end{equation}

With a macroscopic shear force of the form $F(t)/W=\gamma(t)=\gamma_{dc}+\gamma_{ac} 
\cos\left(\omega t\right)$, we obtain for the displacement response:
\begin{equation}
    \label{eq:M7}
{ 
       x_{\searrow}=-\mu\lambda\ln\left(1-\frac{\gamma_{max}}{\mu}
       \right)+2\mu \lambda \ln 
       \left[1+\frac{\gamma_{ac}\left(\cos (\omega 
       t)-1\right)}{2\mu} \right]
       }
\end{equation}
and
\begin{equation}
    \label{eq:M8}
{
       x_{\nearrow}=-\mu\lambda\ln\left(1-\frac{\gamma_{max}}{\mu}
       \right)+2\mu\lambda
       \ln\left(1-\frac{\gamma_{ac}}{\mu}\right)-2\mu \lambda \ln 
       \left[1-\frac{\gamma_{ac}\left(\cos (\omega 
       t)+1\right)}{2\mu} \right]
       }
\end{equation}

\subsection{In-phase and out-of-phase components}

The elastic and dissipative responses are given by:
\begin{equation}
x_{0}=\frac{2}{T}\int_{0}^{\frac{2\pi}{\omega}}x \cos (\omega t)dt
=\frac{\omega}{\pi}\left\{ 
\int_{0}^{\frac{\pi}{\omega}}x_{\searrow}\cos (\omega t)dt + 
\int_{\frac{\pi}{\omega}}^{\frac{2\pi}{\omega}}x_{\nearrow}\cos (\omega t)dt
\right\}
\end{equation}

\begin{equation}
x_{90}
=\frac{\omega}{\pi}\left\{ 
\int_{0}^{\frac{\pi}{\omega}}x_{\searrow}\sin (\omega t)dt + 
\int_{\frac{\pi}{\omega}}^{\frac{2\pi}{\omega}}x_{\nearrow}\sin (\omega t)dt
\right\}
\end{equation}

We expand in power of $\gamma_{ac}/(2\mu)\ll 1$ the time-dependent logarithmic terms in equations (\ref{eq:M7}) and 
(\ref{eq:M8}), and finally get for the in-phase amplitude:
\begin{equation}
    \label{eq:M9}
    {
x_{0}=2\mu\lambda\left[\frac{\gamma_{ac}}{2\mu}+\left(\frac{\gamma_{ac}}{2\mu}\right)^{2}
+\frac{5}{4}\left(\frac{\gamma_{ac}}{2\mu}\right)^{3}+
 \mathcal{O}\left(\frac{\gamma_{ac}}{2\mu}\right)^{4}\right]
          }
\end{equation}

The out-of-phase component can be calculated exactly and reads:

\begin{equation}
    \label{eq:M10}
    {
x_{90}=\frac{4\mu\lambda}{\pi}\left[\left(1-\frac{2\mu}{\gamma_{ac}}\right)\ln\left(
1-\frac{\gamma_{ac}}{\mu}\right)-2\right]
           }
\end{equation}


\end{document}